\documentclass[preprint,showpacs,preprintnumbers,amsmath,amssymb]{revtex4}
\draft

\begin{document}

\title{An inversion formula for transport equation in 3-dimensions using several complex variable analysis}

\author{  S. M. \surname{Saberi Fathi}}

 \altaffiliation{  Department of Physics, Ferdowsi University of Mashhad, Mashhad, IRAN.
 \texttt{email. saberifathi@um.ac.ir.}}

\vspace{1cm}

\begin{abstract}
{\footnotesize
In this paper, the photon stationary transport equation has been extended from $\mathbb{R}^3$ to $\mathbb{C}^3$.
A solution of the inverse problem is obtained on a hyper-sphere and a hyper-cylinder as X-ray and Radon transform, respectively. We show that these results 
can be transformed into each other and they agree with known results.}
\end{abstract}


\maketitle

\section*{Introduction}
The transport equation has many applications in different fields of science and technology (see \cite{golfier}-\cite{chen} and the references therein).
The solution of the stationary non-diffusive transport equation for photons solves many imaging inverse problems
such as Computerized Tomography (CT), Single Photon Emission Computerized Tomography (SPECT) and Positron Emission Tomography (PET) \cite{novikov}-\cite{jpa_non}.

In three dimensions, the solution of transport equation, without attenuation effect, appears as the``X-ray cone beam" transform, without any  restriction on the set of source points $\mathbf{x}$ in \cite{solmon,SmithK,  Hamaker, helgason}. The reconstruction formula is obtained by averaging the X-ray data on the unit sphere in $\mathbb{R}^{3}$. A special case with application to tomography
considers point sources lying on a space curve and is treated in \cite{kirillov, tuy, finch85, palamodov}. B D Smith has introduced an approach for converting divergent beam data into parallel beam data in order to use the well-known Radon inversion procedure \cite{smith}. In another work, P Grangeat has made a conversion of X-ray data into 3-dimensional Radon data in order to apply the Radon inverse formula  \cite{grangeat}.

In two dimensions,  Novikov \cite{novikov} has worked out an explicit inversion formula for transport equation where attenuation effects are taken into account. In his work, Novikov has used an extension $\mathbb{R}^2$ to $\mathbb{C}$ and apply the procedure of analytical continuation. The inversion problem of the transport equation with attenuation in three dimensions has been solved by \cite{jpa_non}, using quaternion analysis. The aim of this paper is to use several-complex analysis to find an inversion formula for the transport equation in $\mathbb{R}^3$. We would also like to use this idea to find an explicit relation between the x-ray transform and Radon transform solutions. We will also compare our results with existing ones in the literature and we show that they are the same.

This paper organized as follows: in the next section we complexify the photon transport equation. Then, in section 2 we obtain the inverse problem solution for the unit ball subspace as x-ray divergent beam transform. In section 3 we use cylindrical coordinate subspace to find the solution. After that we present the results for the three-dimensional case in section 4. In section 5, we will convert the two results into each other. Finally a conclusion and perspectives are given in the last section.

\section{Photon Transport Equation in Several Complex}

Photon transport is considered as non-interacting phenomenon with light speed, $ c $. For simplicity we set $c=1$ and consider the stationary photon transport equation 
\begin{equation}\label{ }
\mathbf{n}\cdot\nabla_{x}\Phi_1(\mathbf{x})= \rho_1(\mathbf{x}),
\end{equation}
where  $\Phi_1(\mathbf{x})$ is the photon flux density, $\rho_1\in \mathcal{C}^1_0$ the photon source term and  $\mathbf{n}\equiv (n_1,n_2,n_3)$ is the photon motion direction. 
We now consider the analytic continuation of this equation in several complex variable space.
In two dimensions, this procedure is known, i.e. $x\longrightarrow x+ i y,\,\,(y\longrightarrow 0)$. In three dimensions, analytic continuation requires 
complexification of the photon direction vector, as done in  \cite{novikov,bal,st}. This  
can be done by adding a second photon transport equation with respect to a variable $ y $ in a perpendicular direction to the initial photon direction, $\textbf{n}$. 
We call this direction $\textbf{n}^\bot\equiv(n_1^\bot,n_2^\bot,n_3^\bot)$. For spatial dimension equal or greater than 3, 
the definition of $\textbf{n}^\bot$ is not unique: $\textbf{n}^\bot$ can be any vector in the perpendicular hyper-plane to $\textbf{n}$ .  
We do not worry about definition of $\textbf{n}^\bot$ because we will see that in the final step we take an average over all possible directions. 
Thus, this procedure leads to the following equation
\begin{equation}\label{3}
\left(\mathbf{n}\cdot\nabla_{x}+\mathbf{n}^\bot\cdot\nabla_{y}\right)\Phi(\mathbf{x},\mathbf{y}),
= \rho(\mathbf{x},\mathbf{y})
\end{equation}
where $\rho(\mathbf{x},\mathbf{y})$ is the source term extending $\rho_1(\mathbf{x})$. The above partial differential equation is defined in an $\mathbb{R}^6\equiv(\mathbb{R}^3\times\mathbb{R}^3)$ Euclidean space of $(\mathbf{x},\mathbf{y})$. We now embed it into $\mathbb{C}^3$. We first define
\begin{equation}
\left\{\begin{array}{l}
\mathbf{r}=\mathbf{x}+\mathrm{i}\,\mathbf{y}\\ \nabla=\nabla_x-\mathrm{i}\,\nabla_y\\ \mathbf{N}=\mathbf{n}+\mathrm{i}\,\mathbf{n}^\bot
\end{array}\right.
\end{equation}
and then we impose the following condition
\begin{equation}\label{31}
\left(\mathbf{n}^\bot\cdot\nabla_{x}-\mathbf{n}\cdot \nabla_{y}\right)\Phi(\mathbf{x},\mathbf{y}) = 0,
\end{equation}
where $\Phi$ is a function defined on a domain in a closed set $\Omega\subset \mathbb{R}^6$, 
this condition means that $\Phi|_{\check{S}}=\phi$, where $ \check{S}=\{z_i\in\mathbb{C}^3,|z_i|=1,i=1,2,3\} $ is the Shilov boundary. The above condition means $\rho$ is a function with compact support on $\partial\Omega$. This condition also means that there is a holomorphic function such as  $\phi\in \Omega$ so that $\overline{\partial} \phi=0$ and $\phi$ has an holomorphical extension on $\Omega$ \cite{Mario}, such as $\Phi$. Now, we can write the equivalent of  (\ref{3}) in $\mathbb{C}^3$ as:
\begin{equation}\label{4}
\left\{\begin{array}{ll}
  \mathbf{N}\cdot \nabla \Phi(\mathbf{x},\mathbf{y})=\rho(\mathbf{x},\mathbf{y}),&(\mathbf{x}, \mathbf{y})\in\Omega\subset\mathbb{R}^6\\     \Phi|_{\check{S}}=0,
     \end{array}\right.
\end{equation}
In fact, the above equation is  a simplified stationary transport equation with source or sink term and without  attenuation \cite{barrett}. Since $\mathbf{N}\cdot\nabla$ is a directional derivative, its inverse is x-ray divergent beam integration\footnote{This is not the second order ultra-hyperbolic partial differential equation of F John \cite{john}.} 
\begin{equation}\label{ }
(\mathcal{X}\widetilde{\rho})(\mathbf{r},\mathbf{N})=\int_{0}^{\infty}dt\,\,\widetilde{\rho}(\mathbf{r}+t\mathbf{N}).
\end{equation}
The physical boundary condition on  $(\mathcal{X}\rho)(\mathbf{r},\mathbf{N})$ is the following. For a given direction $\mathbf{N}$, because of the support hypothesis and the prescription on the direction of integration, $(\mathcal{X}\widetilde{\rho})(\mathbf{r},\mathbf{N})|_{\check{S}}=\phi$, and $\mathbf{N}$ points outward of $\partial\Omega$.

\bigskip
Here $\mathbf{N}$ depends only on three independent parameters because of $\mathbf{n} \cdot\mathbf{n}^\bot=0,~~\frac{|\mathbf{N}|}{\sqrt{2}}=|\mathbf{n}^\bot|=|\mathbf{n}|=1$ and also $\mathrm{i}\,\overline{\mathbf{N}}=\mathbf{N}^\bot$ which yields:  $\textbf{N}\cdot \textbf{N}^\bot=2\mathrm{i}$. 

\bigskip

The x-ray divergence beam transform of  $\rho$ is a superposition of Radon transform in $\mathbf{N}$-direction for \big($(\mathcal{R}\rho)(\mathbf{r},\mathbf{N})=\int_\mathbb{R}dt\,\,\rho(\mathbf{r}+t\mathbf{N})$\big) and in $(-\mathbf{N})$-direction \big($(\mathcal{R}\rho)(\mathbf{r},-\mathbf{N})=\int_\mathbb{R}dt\,\,\rho(\mathbf{r}-t\mathbf{N})$\big),  i.e. x-ray transform of $\rho$ is \cite{bers,st}:
\begin{equation}\label{xR}
(\mathcal{X}\rho)(\mathbf{r},\mathbf{N})=(\mathcal{R}\rho)  (\mathbf{r},\mathbf{N})+(\mathcal{R}\rho)(\mathbf{r},-\mathbf{N})
\end{equation}
In fact, the above relation and the Paley-Wienner's theorem for the Radon transform \cite{gelfand} allowed us to extend compactly supported space $ \Omega$ to a subspace of $\mathbb{C}^3$ with an  holomorphical continuation.  For fully complexified equation (\ref{4}) We have to extend $\mathbf{N}$ in complex form also. 
This extension will be done by generalizing the method of \cite{novikov,st}. Let us define the following parameters and variables:
\begin{equation}\label{cv}
\left\{\begin{array}{ll}
 \lambda_i=|\lambda_i|e^{\mathrm{i}\theta_i}=|\lambda_i|(n_i+\mathrm{i}n_i^{\bot}),  \\ z_i=x_i+\mathrm{i}y_i,\\ \overline{z_i}=x_i-\mathrm{i}y_i,& i=1,2,3.
\end{array}\right.
\end{equation}
where $\theta_i$ is the angle between $x_i$-axis and $\mathbf{x}$ following Lemma gives the holomorphical extension of equation (\ref{3}).

\bigskip

\noindent \textbf{Lemma 1.} \textit{The extended several complex variable form of equation (\ref{3}) is:}
\begin{equation}\label{8}
\nabla_\lambda\Phi(\mathbf{r},\lambda)\equiv
\sum_{i=1}^{3}\frac{\partial\,\widetilde{\Phi_i}(\eta)}{\partial\,\overline{\eta_i}} =\rho(\mathbf{r})
\end{equation}
\textit{where $\eta$ is defined in (\ref{7}).}

\textbf{Proof.} By choosing the parameters and variables in the equation (\ref{cv}), the derivative operator in (\ref{3}) becomes as
\begin{equation}\label{5}
\left(\mathbf{n}\cdot\nabla_{x}+\mathbf{n}^\bot\cdot\nabla_{y}\right)= \frac{1}{2}\,\sum_{i=1}^3\left(\,e^{\mathrm{i}\theta_i}\frac{\partial}{\partial z_i}+\frac{1}{e^{\mathrm{i}\theta_i}}\frac{\partial}{\partial\overline{z}_i}\right).
\end{equation}
Extending the above equation to complex, $\lambda_i$, yields
\begin{eqnarray}\label{6}
\nonumber \left(\mathbf{n}\cdot\nabla_{x}+\mathbf{n}^\bot\cdot\nabla_{y}\right)\rightarrow \nabla_\lambda= \sum_{i=1}^N\,\nabla_{\lambda_i} &=& \frac{1}{2}\,\sum_{i=1}^N\left(|\lambda_i|\,e^{\mathrm{i}\theta_i}\frac{\partial}{\partial z_i}+\frac{1}{|\lambda_i|\,e^{\mathrm{i}\theta_i}}\frac{\partial}{\partial \overline{z}_i}\right)\\ &=&\frac{1}{2}\,\sum_{i=1}^N\left(\lambda_i\frac{\partial}{\partial z_i}+\frac{1}{\lambda_i}\frac{\partial}{\partial\overline{z}_i}\right).
\end{eqnarray}
Here  $\nabla_\lambda$ is the several complex form of left-hand side of (\ref{5}) and when $|\lambda_i|\rightarrow1$ then, $\nabla_\lambda\rightarrow$
`left-hand side of (\ref{5})'. Now consider the following change of variables
\begin{equation}\label{7}
 \eta_i=\frac{1}{\mathrm{i}}\left(\lambda_i^{-1} \,z_i -
\lambda_i\,\overline{z}_i\right).
\end{equation}
Setting $\eta_i=\Re(\eta_i)+\mathrm{i}\Im(\eta_i)$, we have
\begin{equation}\label{ }
\left.\begin{array}{ll}
 \Re(\eta_i)=(|\lambda_i|+|\lambda_i|^{-1})s_i,& \Im(\eta_i)=(|\lambda_i|-|\lambda_i|^{-1})t_i,
\end{array}\right.
\end{equation}
where
 \begin{equation}\label{ }
\left\{\begin{array}{ll}
t_i=(\mathbf{r}\cdot\mathbf{N})_i=x_i\cos\theta_i+y_i\sin\theta_i,\\s_i=(\mathbf{r}\cdot\mathbf{N}^\bot)_i=-x_i\sin\theta_i+y_i\cos\theta_i, & (i=1,2,3)
\end{array}\right.
\end{equation}
where  $\mathbf{r}\equiv(\mathbf{x},\mathbf{y})\equiv(x_1,x_2,x_3,y_1,y_2,y_3)$ is a vector in $\mathbb{R}^{6}$, $\zeta\in \mathbb{C}^3$, $\mathbf{t}\equiv(t_1,t_2,t_3)$ and $\mathbf{s}\equiv(s_1,s_2,s_3)$. We can write $\mathbf{r}=\sum_{i=1}^N(N_i\,t_i+N_i^\bot\,s_i)$. We also note that $\mathbf{x}=\sum_{i=1}^{3}\, \,n_i\,x_i $ and $\mathbf{y}=\sum_{i=1}^{3}\, \,n_i^\bot\,y_i $, so that we have $\mathbf{x}\cdot\mathbf{n}^\bot=0$ and $\mathbf{y}\cdot\mathbf{n}=0$.  Using the change of variables introduced in (\ref{7}), the $\nabla_{\lambda_{i}}$ of equation (\ref{6}) becomes
\begin{equation}\label{15}
\nabla_{\lambda_{i}}=\frac{1}{2}\,\left(\lambda_i\frac{\partial}{\partial z_i}+\frac{1}{\lambda_i}\frac{\partial}{\partial\overline{z}_i}\right)= \frac{1}{2\mathrm{i}}\left(|\lambda_i|^{2} - |\lambda_i|^{-2}\right)\,\frac{\partial}{\partial \overline{\eta_i}}.
\end{equation}
Defining now the function
\begin{equation}\label{1_6}
\widetilde{\Phi_i}(\eta)=\widetilde{\Phi_i}(\Re(\eta),\Im(\eta))=\frac{1}{2\mathrm{i}}\,\left(|\lambda_i|^{2} - |\lambda_i|^{-2}\right)\, \Phi_i(\mathbf{r},\lambda),
\end{equation}
we see that
\begin{equation}
\nabla_{\lambda_i}\,\Phi_i(\mathbf{r},\lambda)=\,\frac{\partial \widetilde{\Phi_i}(\eta)}{\partial \overline{\eta_i}}.
\end{equation}
Hence the several complex variable extended form of (\ref{3}) is obtained as  equation (\ref{8}).$\blacktriangle$

\section{Hyper-Spherical Coordinates}
To solve equation (\ref{8}), we shall use the following theorem in the several complex analysis.

\bigskip

\noindent
\textbf{Theorem 2 } \cite{boas,thie}.\emph{ If $h(z)=\sum_{i=1}^3h_i(\eta)d\overline{\eta}_i$ is a compactly supported (0,1)-form in $\mathbb{C}^N$ of class $\mathcal{C}^1$ that is $\overline{\partial}$-closed. Then, the solution of this problem is known as:}
\begin{equation}\label{ }
f(z)=\int_{\mathbb{C}^N} h(\xi)\wedge B(\xi,z)
\end{equation}
\emph{then $f$ is a compactly supported function such that $\overline{\partial}f=h$.  Here $B(\xi,z)$ is the Bochner-Martinelli kernel which is defined as follows:}
\begin{equation}\label{BM}
B(\xi,z) =\frac{(N-1)!}{(2\pi \mathrm{i})^N}\sum_{i=1}^N (-1)^{i+1}\frac{(\overline{\xi_i}-\overline{z_i})}{|\xi-z|^{2N}} \, d\varpi_i(\xi)
=\sum_{i=1}^N (-1)^{i+1}\,G_i(\xi,z)\, \,d\varpi_i(\xi),
\end{equation}
\emph{where}
\begin{equation}\label{ }
   d\varpi_i(\xi)=d\overline{\xi_1}\wedge \cdots \wedge d\overline{\xi_{i-1}}\wedge d\overline{\xi_{i+1}}\wedge\cdots \wedge d\overline{\xi_N}\wedge d \xi_1 \cdots \wedge d \xi_N. \blacktriangle
\end{equation}

\bigskip

We rewrite now (\ref{8}) in $\overline{\partial}\Phi_c=\sum_{i=1}^{3}\,\frac{\partial\widetilde{\Phi_i}}{\partial\overline{\eta_i}}\,d\eta_i$ form as follows
\begin{equation}\label{dbar}
\overline{\partial}\Phi_c(\eta)=h(\eta)
\end{equation}
where $h(\eta)=\sum_{i=1}^3\, \widetilde{\rho_i}(\eta)\,d\overline{\eta}_i$. Thus, in equation(\ref{dbar}) $h$ is a one-form function with compact support in $\Omega\subset \mathbb{C}^3$. $\widetilde{\rho_i}(\eta)$ is the several complex variable form of $\rho$ and in our case $\widetilde{\rho_i}(\eta)$ is the same for all $i$. We should add that $\rho$ (equivalently $\widetilde{\rho_i}(\eta)$) is a holomorphic function, i.e. $\frac{\partial\,\widetilde{\rho_i}(\eta)}{\partial\overline{\eta_i}}=0, \forall i\in\{i|i=1,2,3\}$, thus, it satisfies the integrability condition in several complex variable analysis $\frac{\partial\widetilde{\rho_i}}{\partial\eta_j}=\frac{\partial\widetilde{\rho_j}}{\partial\eta_i}$. Now, we  use the above theorem to obtain the solution of our system of equations. But we must prove first the following Lemma.

\bigskip

\noindent\textbf{Lemma 3.}\textit{ The solution of the inverse problem equation  (\ref{8}) is: }
\begin{equation}\label{sp}
\rho(\mathbf{r})=\,\frac{1}{2\,\mathrm{i}\,\pi^3} \sum_{i=1}^3\, \int \,N_i\,\frac{\partial}{\partial\overline{z_i}} (\mathcal{X}\rho)(\mathbf{r},\mathbf{N})\, d\Gamma
\end{equation}
\textit{with} $N_i\,\frac{\partial}{\partial\overline{z_i}} =(n_i+\mathrm{i}n_i^\bot)\left(\frac{\partial}{\partial x_i}+\mathrm{i}\frac{\partial}{\partial y_i}\right)$.

\bigskip

\textbf{Proof.} By using Theorem 2 we can write $\widetilde{\Phi_i}$ as:
\begin{equation}\label{phi1}
\widetilde{\Phi_i}(\eta)=\,\int_\Omega\,  \left(|\lambda_i|^{2} - |\lambda_i|^{-2}\right)^{-1}\, \widetilde{\rho_i}(\xi)\,G_i(\xi-\eta) \, d\xi
\end{equation}
where $d\xi$ indicates the integration volume element defined as
\begin{equation}\label{ }
d\xi=(-1)^{i+1}\,d\overline{\xi_i}\,\wedge\,d\varpi_i(\xi)  = d\overline{\xi_1}\wedge d\xi_1\cdots\wedge d\xi_3.
\end{equation}
Then $\widetilde{\Phi_i}$ in equation (\ref{phi1}) can be written as:
\begin{equation}\label{18}
\widetilde{\Phi_i}(\eta)=\frac{1}{4\pi^3 }\int_\Omega\, \frac{\overline{\xi_i}-\overline{\eta_i} } {|\xi-\eta|^{6}}\,  \, \widetilde{\rho_i}(\xi)\,\,d\xi.
\end{equation}
 Setting $|\lambda|\rightarrow0$ and  making the following change of variables  $(\overline{\xi_i}-\overline{\eta_i})\rightarrow \eta_i = \lambda_i^{-1}\,z_i=\lambda_i^{-1}(\mathbf{r}\cdot \mathbf{N})_i=\lambda_i^{-1}\,N_i\,t$ means that we have considered a source in $\xi$ with photon flying in the $\mathbf{N}-$direction. This means that $r_i\rightarrow r_i+\,N_i\,t$.  Thus, by using equation (\ref{1_6}) the above formula yields $\Phi(\mathbf{r},\mathbf{N})$ as follows
\begin{equation}\label{ }
\Phi(\mathbf{r},\mathbf{N})=\,\frac{ 1}{\mathrm{i}\,\pi^3 }\,\sum_{i=1}^3\, \int_\Omega\, \frac{\lambda_i^{-1}\,N_i\,t } {2^3\,|\lambda|^{-6}\,t^6}\, \frac{\rho(\mathbf{r}+\mathbf{N}\,t)}{|\lambda_i|^{-2}}\,d\xi, \,\,\,   \, \left(|\lambda|^2=\sum_{i=1}^3 |\lambda_i|^2\right).
\end{equation}
Using $|\lambda_i|\,N_i=\lambda_i$ and spherical volume element for $d\xi=2^3\,|\lambda|^{-6}\,t^5\,dt\,d\Gamma, ~t\in\mathbb{R}^+$, we obtain
\begin{equation}\label{ }
\Phi(\mathbf{r},\mathbf{N})=\,\frac{1 }{\mathrm{i}\,\pi^3 }\sum_{i=1}^3\,\int_\Gamma\, \lambda_i\,N_i\, (\mathcal{X}\rho)(\mathbf{r},\mathbf{N})\,d\Gamma
\end{equation}
For $|\lambda_i|=0$, we have also $\nabla_{\lambda}\,=\, \frac{1}{2}\,\sum_{i=1}^3\,\lambda_i^{-1}\frac{\partial}{\partial \overline{z}_i}$ and equation (\ref{sp}) is proved.$\blacktriangle$

\section{Hyper-Cylindrical Coordinates}
To obtain the solution of the equation (\ref{8}) in the cylindrical coordinates we shall use the following theorem:

\bigskip

\noindent
\textbf{Theorem 4} \cite{blaya}. \emph{The fundamental  solution of the following problem in  $\mathbb{C}^N$ }
\begin{equation}\label{ }
\frac{  \partial}{\partial  \overline{z}}G(z)=\delta(\mathbf{x},\mathbf{y}), ~~~~z(=\mathbf{x}+\mathrm{i}\mathbf{y})\in\mathbb{C}^N , \mathbf{x},\mathbf{y}\in\mathbb{R}^N
\end{equation}
\emph{is:}
\begin{equation}\label{11}
G(z)=\sum_{i=1}^N\,G_i(z)=\,-\,\frac{(N-1)!}{(2\pi \mathrm{i})^N}\sum_{i=1}^N \frac{\overline{z_i}}{|z|^{2N}}.\blacktriangle
\end{equation}

\bigskip
We prove first the lemma.

\bigskip
\noindent\textbf{Lemma 5.}\textit{ The solution of the equation}
\begin{equation}
\nabla_{\lambda}\widetilde{G}(\eta)\,=\, \delta(\mathbf{t},\mathbf{s}),~~~(i=1,2,3)
\end{equation}
\textit{ is:}
\begin{equation}\label{Gi}
\widetilde{G}(\eta)=\sum_{i=1}^{3} \widetilde{G_i}(\eta), ~~~\widetilde{G_i}(\eta)=\,-\,
\frac{\mathrm{sgn}(|\lambda_i|^2-|\lambda_i|^{-2})}{2\,\pi^3 }  \,\,  \frac{\overline{\eta_i}}{|\eta|^{6}}.
\end{equation}

\bigskip

\textbf{Proof.} We start with $\nabla_{\lambda}\widetilde{G}(\eta)=\delta(\Re(\eta),\Im(\eta)) ,~~\eta\in\mathbb{C}^3$ and use equation (\ref{15}). Hence
\begin{eqnarray}\label{ }
\nonumber\frac{1}{2\,\mathrm{i}}\,(|\lambda_i|^2-|\lambda_i|^{-2})\,  \frac{\partial}{\partial\overline{\eta_i}}\left(
\frac{-1}{4\pi^3 \mathrm{i}}\frac{\overline{\eta_i}}{|\eta|^{6}}\right) &=&\delta(\Re(\eta_i),\Im(\eta_i)) \\ &=&2^{-2}\,|\,|\lambda_i|^{2} - |\lambda_i|^{-2}\,|\,\,\delta(\mathbf{s},\mathbf{t}),
\end{eqnarray}
which, after some algebraic steps, leads to the equation (\ref{Gi}).$ \blacktriangle $

\bigskip

For $|\lambda_i|= (1\pm\epsilon),\, \, \epsilon\rightarrow 0^+$, we have
\begin{equation}\label{Gpm}
\widetilde{G_i^\pm}(\eta)=
\frac{\mp \,1}{2\,\pi^3 }  \,\,  \frac{N_i^\bot}{2^5\,s^{4}\,(s_i\,\pm\mathrm{i} \, \epsilon\,t_i)}=\mp\,\frac{ 1}{2^6\,\pi^3 }  \,\left(\,\mathrm{P.V.}  \frac{1}{s^{4}\,s_i} \mp \, \mathrm{i}\,\pi\,\mathrm{sgn}(t_i) \, \delta(s_i)\,\right).
\end{equation}
Now, the Green function introduced in the above equation allows us to construct the limiting value $\Phi$ of the  solution of
\begin{equation}\label{cy1}
\nabla_\lambda\, \Phi(\mathbf{r}, \mathbf{\lambda})=\rho(\mathbf{r})
\end{equation}

\bigskip

\noindent\textbf{Theorem 6.} \textit{The solution of the inverse problem for the equation (\ref{cy1}) is obtained as}
\begin{equation}\label{CYL}
\rho(\mathbf{r})=\,-\,
 \frac{1}{2\pi\mathrm{i}}\, \sum_{i=1}^3\,  \int \,N_i^{-1}\, \frac{\partial}{\partial\overline{z_i}}\, (\mathcal{H}_i\mathcal{R}\rho)(\textbf{r},\textbf{N}) \, d\Gamma.
\end{equation}

\bigskip

\textbf{Proof.}
The solution $\Phi_i$ is given by
\begin{equation}
\Phi_i(\mathbf{r}, \mathbf{\lambda})=\,\int_\Omega  \,\widetilde{G_i}(\mathbf{r-r'},\mathbf{N})\,\rho(\mathbf{r'})\,dV'=\,\int_\Omega  \,\widetilde{G_i}(\mathbf{r'},\mathbf{N})\,\rho(\mathbf{r-r'})\,dV'.
\end{equation}
The two solutions for $\lambda\rightarrow 1\pm0^+$ becomes $\Phi_i^\pm$ as:
\begin{equation}
\Phi_i^\pm(\mathbf{r}, \mathbf{N})=\,\int_\Omega  \,\widetilde{G_i^\pm}(\mathbf{r'},\mathbf{N})\,\rho(\mathbf{r-r'})\,dV'
\end{equation}
Substituting $\widetilde{G_i^\pm}$ and $dV'=2^6\,dt'\,s^{'4}\,ds' d\Gamma_5$ as hyper-cylindrical coordinates ($d\Gamma_5$ is the spherical surface element of 5-sphere)  we obtain
\begin{equation}
\Phi_i^\pm(\mathbf{r}, \mathbf{N})=\,\mp \,\frac{ 1}{\,\pi^3 }\,\int_{\Gamma_5} (\mathcal{H}_i\mathcal{R}\rho)(\mathbf{r},\mathbf{N})\,d\Gamma_5\,+\,\frac{\mathrm{i}}{\pi^2}\,\int_{\Gamma_5}\,\int_{\mathbb{R}}\,\rho(\mathbf{r}-\mathbf{N}t') \,\mathrm{sgn}(t') \,dt'\,d\Gamma_5.
\end{equation}
Using $\phi=\Phi^+-\Phi^-$ we obtain
\begin{equation}
\phi_i(\mathbf{r}, \mathbf{N})=\,-\,\frac{ 2}{\pi^3 }\,\langle (\mathcal{H}_i\mathcal{R}\rho)(\mathbf{r},\mathbf{N})\rangle_5
\end{equation}
where $\langle[\cdot]\rangle_k=\int_{\Gamma_k}\,[\cdot]\, d\Gamma_k$ indicates average on the k-sphere (in our notation we eliminate subscript for $k=6$ such as $\Gamma\equiv\Gamma_6$), and $\mathcal{H}_i$ is the Hilbert transform over $s_i$ and is defined as
\begin{equation}\label{ }
\mathcal{H}_if :=\frac{1}{\pi}\textsf{P.V.}\int\frac{f(x)}{x_i}dx_i
\end{equation}
where `$\textsf{P.V.}$' indicates the principal value of a singular integral. Finally, this result obtained in cylindric coordinates should be consider as in a cylinder of $\mathcal{C}^3$.  Using Cauchy integral on the variable, which is not on the unit ball (to project on Shilov boundary)  Lemma on page 32 of \cite{Mario} gives
\begin{equation}\label{CI}
\Phi(\eta,\lambda)=\,\frac{1}{2\pi^2\,\mathrm{i}}\,\sum_{i=1}^3\, \int_{|\mu_i|=1} \,\frac{\phi_i(\eta,\mu)}{\mu_i-\lambda_i} d\mu_i,~~~ (\lambda\in\Omega\setminus \Gamma)
\end{equation}
where $\Phi$ is a holomorphic extension of $\phi=\Phi^+-\Phi^-$.
For $\lambda_i\rightarrow 0$ we have $\nabla_\lambda=\frac{1}{2}\sum_{i=1}^3\,\lambda_i^{-1}\,\frac{\partial}{\partial\overline{z_i}}$ and $\frac{1}{\mu_i-\lambda_i} =\frac{1}{\mu_i}\left(1+\frac{\lambda_i}{\mu_i}+\cdots\right)$. Thus,
the following formulas hold
\begin{eqnarray}
\rho(\textbf{r})&=& \,-\,\frac{1}{4\pi^2\,\mathrm{i}}\,\sum_{i=1}^3\, \frac{\partial}{\partial\overline{z_i}}\, \int_{|\mu_i|=1} \,\frac{\phi_i(\textbf{t},\textbf{s})}{\mu_i^2} d\mu_i \\ &=&\,-\, \frac{1}{4\,\pi^2\,\mathrm{i}}\, \sum_{i=1}^3\, \frac{\partial}{\partial\overline{z_i}}\, \int_{|\mu_i|=1} \,\mu_i^{-1}\,\langle \mathcal{H}_i\mathcal{R}\rho (\textbf{t},\textbf{s})\rangle_5 \,d\theta_i,~~~ (\lambda\in\Omega\setminus \Gamma)
\end{eqnarray}
On the unit ball `$|\mu_i|= 1$', it gives equation (\ref{CYL}). $ \blacktriangle $

\section{Three-dimensional results} When $\mathbf{y}\rightarrow 0$, $g(\mathbf{r})\rightarrow g(\mathbf{x})$ then $\nabla_y g(\mathbf{x})=0$  and $d\Gamma\rightarrow 4\pi d\Gamma_3$ where $\Gamma_3$ is the surface of a 3-sphere and by using (\ref{31}), we also have $,N_i^{-1}\, \frac{\partial}{\partial\overline{z_i}}\longrightarrow \mathbf{n}\cdot \nabla_x$, Thus,
\begin{equation}\label{ }
\rho(\mathbf{x})=\frac{1}{2\,\pi^3}\, \int_{\Gamma_3}\, \mathbf{n}\cdot \nabla_x\,
\left\{\begin{array}{l}
\,-\,2\,\pi\,\sum_{i=1}^3\,(\mathcal{H}_i\mathcal{R}\rho)(\textbf{x},\textbf{n})\\ \,(\mathcal{X}\rho)(\mathbf{x},\textbf{n})
\end{array}\right.
\, d\Gamma_3, ~~~(\mathbf{x}\in\mathbb{R}^3)
\end{equation}
This equation includes the x-ray transform, a result obtained by \cite{jpa_non}. 
Moreover we can also show that the term containing the Radon transform is equal to the other term.

\bigskip

\textbf{Remark 7.} $\widetilde{g}=\langle \mathcal{H}_i\mathcal{R}\rho\rangle_5$ is an average on a $5-$sphere, while $g=\langle \mathcal{X}\rho \rangle$ is an average over a $6-$sphere. In reality, this lack of averaging over an angle in the cylinder is compensated by an  holomorphical extension via Cauchy integral.  In analogy with potential theory, the Cauchy kernel is the same as $r^{-1}$ potential, which is a central field potential in physics and it describes circular motion of a particle. On the other hand, the Cauchy integral in equation (\ref{CI}) averages over free angles and completes the averaging over a $6-$sphere. This means that the $6-$cylinder is topologically converted to a $6-$sphere. In tomography, this procedure fulfills Tuy's x-ray inversion formula condition, which states the ``For any bounded object, a curve consisting of two circles [...] or some type of spiral curve will satisfy all the conditions on the curve required in the theorem."\cite{tuy} In three dimensions, Tuy's condition is fulfilled by a sphere because great circles exist on a sphere.

\section{Connection Between Two Approaches}
We also verify the inter-convertibility these results by using the following proposition:

\bigskip

\noindent\textbf{Proposition 8} \cite{finch87}.\textit{ Suppose that $f\in \mathcal{C}^1_0$ and $\mathbf{x}\in\mathbb{R}^3$ and $\theta\in\mathbb{S}^2$. Then}
\begin{equation}
\int_{\Gamma_3} \mathcal{X}f(\mathbf{x},\mathbf{n}_1) \delta^{'}(\mathbf{n}\cdot\mathbf{n}_1) d\Gamma_3=-(\mathcal{R}f)'(\mathbf{x}\cdot\mathbf{n}).\blacktriangle
\end{equation}

\bigskip

\noindent\textbf{Lemma 9.} \textit{Using the above proposition we have} $ \langle\mathcal{X}\rho\rangle=\,- \,2\,\pi\,\sum_{i=1}^{3}\langle \mathcal{H}_i\mathcal{R}\rho\rangle $.

\bigskip

\textbf{Proof.} Here some work on the left-hand side of this proposition yields
\begin{eqnarray}
\nonumber \int_{\Gamma_3} \,\mathcal{X}f(\mathbf{x},\mathbf{n}_1)\, \delta^{'}(\mathbf{n}\cdot\mathbf{n}_1)\, d\Gamma_3&=&- \int_{\Gamma_3} \,(\mathcal{X}f)^{'}(\mathbf{x},\mathbf{n}_1)\, \delta(\mathbf{n}\cdot\mathbf{n}_1)\, d\Gamma_3\\ \label{gel}  &=&-\,\left(\,\int_{\Gamma_3} \dfrac{(\mathcal{X}f)^{'}(\mathbf{x},\mathbf{n}_1)}{ (\mathbf{n}\cdot\mathbf{n}_1)+\,\mathrm{i}\, 0}\,+ \int_{\Gamma_3} \dfrac{(\mathcal{X}f)^{'}(\mathbf{x},\mathbf{n}_1)}{ (\mathbf{n}\cdot\mathbf{n}_1)-\,\mathrm{i}\, 0} d\Gamma_3\right)\\  &=& \,-\,2\,\pi\,\mathcal{H}(\mathcal{X}f)^{'}(\mathbf{x},\mathbf{n}),
\end{eqnarray}
where `$(\cdot)^{'}$' is the differentiation with respect to $\mathbf{n_1}$. In equation(\ref{gel}), we have used the definition of $\delta-$function as a generalized function \cite{gelfand1}. Note that in $\mathcal{H}(\mathcal{X}f)^{'}$ the Hilbert transform  is relative to $\mathbf{n}$. Equating this result to the right-hand side of the equation in the previous proposition, we obtain $2\,\pi\,\mathcal{H}(\mathcal{X}f)^{'}(\mathbf{x},\mathbf{n})=(\mathcal{R}f)^{'}(\mathbf{x}\cdot\mathbf{n})$. By taking another Hilbert transform on both sides of this equation and using `$\mathcal{H}^2=-1$' we obtain   $-2\,\pi\,(\mathcal{X}f)^{'}(\mathbf{x},\mathbf{n})=\mathcal{H}(\mathcal{R}f)^{'}(\mathbf{x}\cdot\mathbf{n})$. These two last relations  between the x-ray divergent beam transform and the Radon transform, equation (\ref{xR}) imply that $-2\,\pi\,(\mathcal{X}f)(\mathbf{x},\mathbf{n})=(\mathcal{H}\mathcal{R}f)(\mathbf{x}\cdot\mathbf{n})$.  The following equation holds: $(\mathcal{H}\mathcal{R}f)(\mathbf{x}\cdot\mathbf{n})=\sum_{i=1}^{3}(\mathcal{H}_i\mathcal{R}f)(\mathbf{x},\mathbf{n})$, where $\mathcal{H}_i$ is Hilbert transform relative to $s_i$. This relation finalizes our proof. $ \blacktriangle $

\bigskip

\section*{Conclusion}
A solution of the transport equation in three dimensions without attenuation effect is obtained by using its extension to the several complex variable space $\mathbb{C}^3$. We have shown that the solution of this extended transport equation in two different subspaces with cylindrical and with spherical coordinates could give rise to two different approaches of solution in terms of Radon and x-ray divergent beam transforms respectively. However, the two different forms of these solutions are convertible into each other. This paper also is an application of several complex variable analysis and its usefulness. In future investigations, one may seek the solution of the inverse problem for transport equation with attenuation using this several complex variable analysis.



\begin{thebibliography}{0}
\expandafter\ifx\csname natexlab\endcsname\relax\def\natexlab#1{#1}\fi
\expandafter\ifx\csname bibnamefont\endcsname\relax
  \def\bibnamefont#1{#1}\fi
\expandafter\ifx\csname bibfnamefont\endcsname\relax
  \def\bibfnamefont#1{#1}\fi
\expandafter\ifx\csname citenamefont\endcsname\relax
  \def\citenamefont#1{#1}\fi
\expandafter\ifx\csname url\endcsname\relax
  \def\url#1{\texttt{#1}}\fi
\expandafter\ifx\csname urlprefix\endcsname\relax\def\urlprefix{URL }\fi
\providecommand{\bibinfo}[2]{#2}
\providecommand{\eprint}[2][]{\url{#2}}

\end{thebibliography}


\begin{thebibliography}{99}
\begin{footnotesize}

\bibitem{golfier} F. Golfier, M. Quintard, S. Whitaker, Journal of Porous Media,  \textbf{5} Issue 3 2002

\bibitem{} S-M Hong, A-T Pham, Ch Jungemann, \emph{Deterministic Solvers for the Boltzmann Transport Equation}, (Springer-Verlag, Wien 2011)

\bibitem{quintard} Michel Quintard, S. Whitaker, Advances in Water Resources, \textbf{17} Issue 4, 1994, Pages 221-–239

\bibitem{zanotti}  F. Zanotti, R.G. Carbonell, Chemical Engineering Science, \textbf{39} Issue 2, 1984, Pages 299-–311

\bibitem{chandr} P. Chandramohan, B.U. Nayak, V.S. Raju,  Coastal Engineering, 12, Issue 3, September 1988, Pages 285-–297

\bibitem{ianna} G. Iannaccone, A. Betti, G. Fiori,   IEEE Conference Publications,  6-8 Sept. 2010, 3-6

\bibitem{chen} Z. Chen, G. Huan, Y. Ma, \emph{Computational Methods for Multiphase Flows in Porous Media}, SIAM 2006

\bibitem{novikov} R. G. Novikov, Ark. Mat. \textbf{40}, 2002, 145-167.

\bibitem{bal} G. Bal  Columbia university Lecture Notes in http://www.colombia.edu/?gb2030/ 2004

\bibitem{jpa_non} S. M. Saberi Fathi,  J. Phys. A: Math. Theor. \textbf{43} (2010) 335202

\bibitem{solmon} D. C. Solmon \emph{Jour. Math. Anal. Appl.} \textbf{56}, 1976, 61-83.

\bibitem{SmithK} k. T.  Smith, D. C. Solmon and S. L. Wagner  \emph{Bulletin of the American Mathematical Society }\textbf{83}(6), 1977, 1227-1270.

\bibitem{Hamaker} C.Hamaker, K. T. Smith, d. C. Solmon D C, and S. L. Wagner \emph{Rocky Mount. J. Math}., \textbf{10}(1), 1980, 253-283.

\bibitem{tuy} H. K.  Tuy  SIAM \emph{J. Appl. Math}. \textbf{43}, 1983 546-52.

\bibitem{helgason} S. Helgason  \emph{The Radon Transform} (Birkh\"{a}user, Berlin, 1999).

\bibitem{kirillov} A, A. Kirillov, \emph{Dokl. Akad. Nauk SSSR} \textbf{137}, 276-277; Eng. Trans. \emph{Soviet Math. Dokl}., \textbf{2}, 1961, 268-269.



\bibitem{finch85} D. V. Finch, \emph{SIAM J. Appl. Math.} \textbf{45}, 1985, 665-673.

\bibitem{palamodov} V. P. Palamodov,  Inversion formulas for the three-dimensional ray transform in \emph{Mathematical Problems of Tomography}, Lecture Notes in Mathematics \textbf{1497}, Eds. G T Herman, A K Louis, F. Natterer, (Springer, Berlin, 1991) 53-62.

\bibitem{smith} B. D. Smith, \emph{Optical Engineering} \textbf{29}(5), 1990, 524-534.

\bibitem{grangeat} P. Grangeat,    \emph{Analysis d'un syst\`eme d'imagerie 3D par r\'econstruction \`a partir de $X$ en g\'eom\'etrie conique}, PhD Thesis, Ecole Nationale Sup\'erieure des T\'e1\'ecommunications, Paris, 1987.




\bibitem{st} S. M. Saberi Fathi  and T. T. Truong  \emph{J. of Phys. A.} \textbf{42} 2009, N° 4

\bibitem{Mario} Mario Landucci, Rendiconi del Seminaro Matematico delta Universit\`{a} di Padova, tome 62 1980, p.23-34.

\bibitem{barrett} H. H. Barrett and W. Swindell,  \emph{Radiological Imaging} I and II, ( Academic Press, New York, 1981)

\bibitem{gelfand} I. M. Gel'fand, M. I. Graev, and N. Ya. Vilenkin, \emph{Generalized Function, Vol 5, Integral Geometry and Representation Theory}, Translated in English by Eugene Saletan, (Academic Press, New York, 1966)

\bibitem{john} F. John  \emph{Plane Waves and Spherical Means Applied to Partial Differential Equations} (Wiley-Interscience, New York, 1955).

\bibitem{boas} H. Boas, Topics in Several Complex Variables, lecture notes, 2005, http://www.math.tamu.edu /~boas/courses/685-2005c/.

\bibitem{bers} S. Bernstein, R. Hielscher and H. Schaeben, Math. Meth. Appl. Sci. \textbf{32} 2009; 379–394

\bibitem{thie} Ch. Laurent-Thi\'ebaut, Transformation de Bochner-Martinelli dans une vari\'et\'e de Stein, Lecture Notes in Mathematics  \textbf{1295}, 1987, pp 96-131

\bibitem{blaya} R. A. Blaya, J. B. Reyes,  Aduvances in Applied Clifford Algebras \textbf{11} No. 1, (2001) 15-26

\bibitem{finch87} D. V. Finch   Approximate reconstruction formulae for the cone beam transform I, unpulished Nov. 1987

\bibitem{gelfand1} I. M. Gel'fand and E. G.  M Shilov  \emph{Generalized Function, Vol 1, Properties and Operations}, Translated in English by Eugene Saletan, (Academic Press, New York, 1964)











%


\end{footnotesize}
\end{thebibliography}
\end{document}